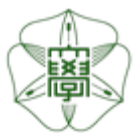

# HOKKAIDO UNIVERSITY

| Title | A neuroeconomic theory of bidirectional synaptic plasticity and addiction |
|---|---|
| Author(s) | Takahashi, Taiki |
| Citation | Medical Hypotheses, 75(4): 356-358 |
| Issue Date | 2010-10 |
| Doc URL | http://hdl.handle.net/2115/44260 |
| Right | |
| Type | article (author version) |
| Additional Information | |

Instructions for use

# A neuroeconomic theory of bidirectional synaptic plasticity and addiction.

Taiki Takahashi[1]

[1] Department of Behavioral Science, Hokkaido University

Corresponding Author: Taiki Takahashi

Email: taikitakahashi@gmail.com

Department of Behavioral Science, Hokkaido University

N.10, W.7, Kita-ku, Sapporo, 060-0810, Japan

TEL: +81-11-706-3057    FAX: +81-11-706-3066

**Acknowledgements:** The research reported in this paper was supported by a grant from the Grant- in-Aid for Scientific Research ("global center of excellence" grant) from the Ministry of Education, Culture, Sports, Science and Technology of Japan.

Summary:

Neuronal mechanisms underlying addiction have been attracting attention in neurobiology, economics, neuropsychiatry, and neuroeconomics. This paper proposes a possible link between economic theory of addiction (Becker and Murphy, 1988) and neurobiological theory of bidirectional synaptic plasticity (Bienenstock, Cooper, Munro, 1982) based on recent findings in neuroeconomics and neurobiology of addiction. Furthermore, it is suggested that several neurobiological substrates such as cortisol (a stress hormone), NMDA and AMPA receptors/subunits and intracellular calcium in the postsynaptic neurons are critical factors determining parameters in Becker and Murphy's economic theory of addiction. Future directions in the application of the theory to studies in neuroeconomics and neuropsychiatry of addiction and its relation to stress at the molecular level are discussed.

## 1. Introduction:

Addiction has been a major topic in neuroeconomics and neurobiology of synaptic plasticity [1], because a better understanding of neurobiological mechanisms is important for the establishment of medical treatments for addiction. In order to quantitatively model addictive behavior, economists Becker and Murphy proposed an economic theory of addiction [2], which illustrates addiction as a habit-forming learning process. A considerable number of behavioral and neuroeconomic studies utilized this theory to analyze various types of addictive behavior [3,4]. It is to be noted that Becker and Murphy's theory of rational addiction contains several key parameters which have, to date, not been associated with neurobiological substrates, as introduced later. Regarding neurobiological mechanisms underlying addiction, synaptic plasticity in dopaminergic neural circuits plays a pivotal role [1]. Furthermore, in order to model synaptic plasticity in a biophysically plausible manner, Bienenstock, Cooper, and Munro proposed a neurobiological theory of bidirectional synaptic plasticity, which contains a key parameter of the threshold for synaptic plasticity [5]. Although (a) Bienenstock-Cooper-Munro (BCM) theory [5] has been utilized to investigate neurobiological mechanisms underlying addiction, learning and memory and (b) neurobiological processes underlying addiction share common mechanisms with learning and memory, no study to date has associated BCM theory with Becker and Murphy's economic theory of addiction. To unify both theories and suggest some future directions of neuroeconomic studies are the objective of the present study.

 This paper is organized in the following manner. In Section 2, I briefly introduce Becker and Murphy's economic theory of addiction and Bienenstock-Cooper-Munro (BCM) theory of bidirectional synaptic plasticity. In Section 3, I introduce a novel framework combining the economic theory of addiction and BCM theory, and explain neuroeconomic and neurobiological correlates of the parameters in the models. In Section 4, some conclusions from this study and future study directions by utilizing the present neuroeconomic theory, and how to test the present theory experimentally are discussed.

## 2. Economic theory of addiction and neurobiological theory of learning
### 2.1 Becker and Murphy's economic theory of addiction
The rational addiction model of economists Becker and Murphy [2] has become one of the standard tools in the economic analysis of the markets for drugs, alcohol, tobacco (nicotine) and other potentially addictive goods. In Becker-Murphy (BM) theory, the

consumer (addict)'s problem is to maximize

$$\int_0^T U(C(t), A(t), S(t))e^{-rt} dt \tag{1}$$

where A(t) is consumption of the addictive drug (e.g., alcohol and nitotine) at time t, C(t) is consumption of non-addictive goods at t, and S(t), which we shall refer to as "consumption capital", is the stock of addiction or habit, built up as a result of past consumption of A (i.e., the physiological effects of past drug intake including both reinforcement due to drug intake and damages to the body due to the past drug intake). The consumer's temporal horizon is from time 0 to time T, and she exponentially discounts the future rewards at the rate of r. The consumer maximizes equation 1 subject to the "equation of motion" for S:

$$\frac{dS(t)}{dt} = A(t) - \delta S(t) \tag{2}$$

where $0 < \delta < 1$ is the "depreciation rate" which indicates the instantaneous rate of temporal decay of the stock of addiction S. By utilizing this theory, Becker and Murphy predicted that subjects will be addicted to harmful drugs when the subjects' r in equation 1 and $\delta$ in equation 2 are large enough; in other words, when subjects are "myopic" in intertemporal choice and have strong capacity of recovery from damages due to the intake of harmful drugs. Our previous neuroeconomic study demonstrated that daily nicotine intake by habitual smokers were positively associated with temporal discounting (corresponding to r in equation 1), consistent with BM theory [6]. However, to date, neurobiological correlates of the parameters in BM theory have been unclear.

## 2.2 Bienenstock-Cooper-Munro theory of synaptic plasticity

Theoretical physicists Leon N. Cooper and colleagues proposed a modified Hebbian synaptic plasticity model to explain the development of the visual cortex [5]. In Bienenstock-Cooper-Munro (BCM) theory, a change in synaptic weight (synaptic plasticity) is bidirectional and dependent on the threshold for synaptic plasticity. If the (instantaneous) postsynaptic activity is lower than the threshold, the synapse is weakened, while if the postsynaptic activity is higher than the threshold, the synapse is strengthened. Also, the threshold for the synaptic plasticity increases as the previous activity of the neuron increases. Namely, if the instantaneous (transient) postsynaptic plasticity is lower and higher in comparison to the persistent (resting) postsynaptic activity, the synapse is weakened (i.e., long-term depression) and strengthened (long-term potentiation), respectively. Mathematically speaking, the BCM theory is expressed as:

$$y = \sum_i w_i x_i \qquad (3)$$

$$\frac{dw_i}{dt} = y(y - \theta_M)x_i - \varepsilon w_i \qquad (4)$$

$$\theta_M = f(y) \qquad (5)$$

where y is the activity of the postsynaptic neuron, $w_i$ is the $i^{th}$ synaptic weight, $x_i$ is the activity of the $i^{th}$ presynaptic neuron, $\theta_M$ is the threshold for the synaptic plasticity, $\varepsilon$ is the temporal decay rate of the $i^{th}$ synaptic weight. Also, $\theta_M$ is an increasing function f of y (i.e., f '(y)>0). Also, f(y) may be a time-averaged (i.e., non-transient and persistent) postsynaptic activity. Notably, we can see, from equation 4, that if $\theta_M$ is large, the synapse is weakened, while if $\theta_M$ is small, the synapse is strengthened. Neurobiological studies revealed that NMDA (N-methyl-D-aspartate) receptor-induced persistent rise in the intracellular calcium concentrations in the postsynaptic neurons may be positively associated with $\theta_M$ [7]. Furthermore, neurobiological studies reported that stress hormones (e.g., cortisol and corticosterone) increases $\theta_M$, while chronic nicotine treatment normalizes $\theta_M$ [8,9]. It is further to be noted that synaptic plasticity is known to be related to the ratio of NMDA/AMPA receptor expressions at the postsynaptic neurons [10] and an increased proportion of NR2B-containing NMDA receptors at silent synapses in the nucleus accumbens [11].

3. **A neuroeconomic theory of addiction and synaptic plasticity**
As stated earlier, to date, no study has unified Becker-Murphy's economic theory of addiction and Bienenstock-Cooper-Munro's neurobiological theory of synaptic plasticity. To unify these theories is the objective of this section. Let us suppose that equation 5 can be generalized as:

$$\theta_M = f_g(y, S), \qquad (6)$$

where $f_g$ is an increasing function of y in equation 3 but a decreasing function of S in equation 1. Namely,

$$\frac{\partial f_g}{\partial y} > 0 \tag{7}$$

and

$$\frac{\partial f_g}{\partial S} < 0. \tag{8}$$

Equation 4 and 8 indicate that if the "consumption capital" S (in equation 1 and 2) increases (i.e., the threshold $\theta_M$ is decreased) due to chronic intake of addictive drugs, the synapse is strengthened. In other words, addiction may correspond to long-term potentiation due to a decrease in $\theta_M$ which is induced by an increase in S. This may account for the habit-forming properties in addiction due to learning processes at the synaptic level. Actually, intake of addictive drugs such as nicotine facilitates long-term potentiation at the synapses in a dopamine-dependent manner [12]. Furthermore, stress hormones increases $\theta_M$ via persistent (i.e., not transient) rise in postsynaptic calcium levels (corresponding to an increase in y in equation 6) [8,13]. When long-term potentiation is impaired by stress, equation 8 predicts that chronic nicotine treatment decreases $\theta_M$, resulting in the recovery of long-term potentiation, consistent with a recent study [9]. Let us now examine the effect of "depreciation rate" $\delta$ in equation 2 on addiction. If the depreciation rate $\delta$ increases, the consumption capital S decreases more rapidly (from equation 2), resulting in a more rapid increase in $\theta_M$. Because the rapid increase in $\theta_M$ may induce long-term depression and neurodegeneration in the hippocampus (i.e., memory impairment) [8], this may explain why large $\delta$ values may be associated with marked tendencies for addiction at the aim of compensating hippocampal neurodegeneration and memory impairment [8], claimed in Becker and Murphy's theory of addiction. Taken together, these considerations may enable us to state that the unified neuroeconomic theory of addiction by generalized BCM theory which incorporates the parameter of the consumption capital in Becker-Murphy theory can account for most prominent features of addiction at the synaptic levels, as well as at the molecular levels.

4. **Implications for neuroeconomics and neurobiolgy of addiction**

Our present theoretical considerations may make it possible to connects the parameter of "consumption capital" S in Becker and Murphy's economic theory of addiction with neurobiological substrates at the cellular and molecular levels. Therefore, future studies in neuroeconomics can establish economic models of addiction in terms of neurobiological substrates at the molecular level (e.g., expression of NMDA/AMPA receptors and postsynaptic calcium levels). We now denote some implications of the present theoretical framework for neurobiological substrates underlying addiction. First, for cellular neurobiologists who can measure intracellular calcium ($[Ca^{2+}]_i$) with calcium indicators such as fura-2 [13], it can be expected that parameter S in Becker and Murphy's theory may be estimated by measuring the resting intracellular calcium concentrations ($[Ca^{2+}]_i$). Note that a higher resting (i.e., persistent and non-transient) $[Ca^{2+}]_i$ (which is positively related to $\theta_M$) indicates a smaller consumption capital S which is positively associated with the chronic intake of addictive drugs (predicted from equation 8). Also, larger depreciation rate $\delta$ may increase the resting $[Ca^{2+}]_i$ via rapid reduction in S (predicted from equation 2 and 8). It is also predicted that if drug-dependent subjects are exposed to stressors which increases the resting $[Ca^{2+}]_i$ [13], they are likely to more dramatically intake addictive drugs which increases S, resulting in a decrease in the resting $[Ca^{2+}]_i$ (i.e., a reduction in $\theta_M$), because high levels of the resting $[Ca^{2+}]_i$ is neurotoxic [13]. Because S may also be related to expressions of NMDA/AMPA receptor subunits, molecular biologists might be able to estimates S by measuring NMDA/AMPA receptor-related mRNAs.

We finally state some predictions from the present theory. Because, as stated earlier, Becker-Murphy theory states that individuals with larger depreciation rate $\delta$ are more likely to be addicted to harmful drugs, it is expected that individuals with more rapidly changing $\theta_M$ (which may be related to more rapidly changing the resting $[Ca^{2+}]_i$ and NMDA/AMPA receptor expressions in dopaminergic circuits such as the nucleus accumbens) due to chronic intake of addictive drugs, may more easily be addicted to the drugs. This prediction could be tested experimentally in future studies in neuroeconomics and neurobiology of addiction. Interestingly, even harmful addictive drugs may be beneficial for neuronal viability, because an intake of the harmful drugs may reduce the resting $[Ca^{2+}]_i$ via an increase in S. Also, neuropharmacological treatments which temporally stabilize $\theta_M$ may prevent the severe addiction. This possibility should be examined by future neuropharmacological studies.